%% file: main.tex
\documentclass[%
 reprint,
 amsmath,amssymb,
 aps,superscriptaddress,
]{revtex4-1}

\usepackage{graphicx}
\usepackage{dcolumn}
\usepackage{bm}

\usepackage{xargs}                      
\usepackage{multirow}
\usepackage{xcolor}
\usepackage{hyperref}
\usepackage{autofigs8}
\usepackage{siunitx} 
\hypersetup{colorlinks=true,allcolors=blue}

\include{commands}

\begin{document}
\title{\FD\  Squeezing for Advanced LIGO}
\overfullrule 0pt 
\parskip0pt
\hyphenpenalty9999

\author{L.~McCuller}  
\email{lee.mcculler@ligo.org}
\author{C.~Whittle}
\email{chris.whittle@ligo.org}
\affiliation{LIGO, Massachusetts Institute of Technology, Cambridge, MA 02139, USA}
\author{D.~Ganapathy}
\affiliation{LIGO, Massachusetts Institute of Technology, Cambridge, MA 02139, USA}
\author{K. Komori}
\affiliation{LIGO, Massachusetts Institute of Technology, Cambridge, MA 02139, USA}
\author{M.~Tse}
\affiliation{LIGO, Massachusetts Institute of Technology, Cambridge, MA 02139, USA}
\author{A.~Fernandez-Galiana}
\affiliation{LIGO, Massachusetts Institute of Technology, Cambridge, MA 02139, USA}
\author{L.~Barsotti}
\affiliation{LIGO, Massachusetts Institute of Technology, Cambridge, MA 02139, USA}
\author{P.~Fritschel} 
\affiliation{LIGO, Massachusetts Institute of Technology, Cambridge, MA 02139, USA}
\author{M.~MacInnis}
\affiliation{LIGO, Massachusetts Institute of Technology, Cambridge, MA 02139, USA}
\author{F.~Matichard}
\affiliation{LIGO, Massachusetts Institute of Technology, Cambridge, MA 02139, USA}
\author{K.~Mason}
\affiliation{LIGO, Massachusetts Institute of Technology, Cambridge, MA 02139, USA}
\author{N.~Mavalvala} 
\affiliation{LIGO, Massachusetts Institute of Technology, Cambridge, MA 02139, USA}
\author{R.~Mittleman}
\affiliation{LIGO, Massachusetts Institute of Technology, Cambridge, MA 02139, USA}
\author{Haocun~Yu}
\affiliation{LIGO, Massachusetts Institute of Technology, Cambridge, MA 02139, USA}
\author{M.~E.~Zucker}  
\affiliation{LIGO, Massachusetts Institute of Technology, Cambridge, MA 02139, USA}
\affiliation{LIGO, California Institute of Technology, Pasadena, CA 91125, USA}
\author{M.~Evans}
\affiliation{LIGO, Massachusetts Institute of Technology, Cambridge, MA 02139, USA}

\date{\today}

\begin{abstract}

The first detection of gravitational waves by the Laser Interferometer
Gravitational-wave Observatory (LIGO) in 2015 launched the era of gravitational
wave astronomy. The quest for gravitational wave signals from objects that are
fainter or farther away impels technological advances to realize ever more
sensitive detectors. Since 2019, one advanced technique, the injection of
squeezed states of light is being used to improve the shot noise limit to the
sensitivity of the Advanced LIGO detectors, at frequencies above $\sim 50$ Hz. 
Below this frequency, quantum back action, in the form of radiation pressure induced motion of the mirrors, degrades the sensitivity.
To simultaneously reduce shot noise at high frequencies {\it and} quantum radiation
pressure noise at low frequencies requires a quantum noise filter cavity with
low optical losses to rotate the squeezed quadrature as a function of frequency.
We report on the observation of \fd\  squeezed quadrature rotation
with rotation frequency of \SI{30}{Hz}, using a 16 m long filter
cavity. A novel control scheme is developed for this \fd\ 
squeezed vacuum source, and the results presented here demonstrate that a
low-loss filter cavity can achieve the squeezed quadrature rotation necessary
for the next planned upgrade to Advanced LIGO, known as ``A+.''

\end{abstract}

\pacs{Valid PACS appear here}
\maketitle



\section{\label{sec:intro}Introduction}

\psub{Quantum noise limits measurement precision}
Quantum noise imposes a
fundamental limitation on the precision of physical measurements. In
gravitational-wave detectors such as Advanced LIGO~\cite{aLIGOdesign}, it
manifests in two ways: shot noise, caused by quantum fluctuations in the arrival
time of photons detected at the interferometer output; and quantum radiation
pressure noise (QRPN), due to quantum fluctuations in the photon flux impinging
on the interferometer mirrors. Both noises have a common
origin---electromagnetic vacuum states entering the open output port of the
interferometer~\cite{Caves1980a}.

\psub{Advanced LIGO with FID squeezing}
During its third observing run, Advanced
LIGO has been detecting roughly one astrophysical event per week, compared to
one per month in previous observing runs. A significant factor in the
sensitivity improvements that has enabled this increased detection rate is the
injection of squeezed states of light into the output port of the Advanced LIGO
detectors~\cite{tsePRL2019}. Squeezed state injection---or simply
``squeezing''---has also been used in the Advanced Virgo and GEO detectors
\cite{VirgoSqueezing, GEOlongterm}. The squeezed states used redistribute
quantum uncertainty to have a minimum in the phase quadrature, thus reducing
shot noise that dominates above $\sim 50$ Hz. As a consequence of the Heisenberg
uncertainty principle, the increased noise in the amplitude quadrature degrades
the sensitivity of the interferometer at lower frequencies due to radiation
pressure that drives motion of the 40 kg mirrors of the interferometer. This
quantum radiation pressure noise has been observed in the Advanced
LIGO~\cite{LigoQRPN} and Virgo~\cite{VirgoQRPN} detectors under current
operating conditions, and limits the level of squeezing which can be profitably
used.

\psub{FD squeezing and A+} To avoid degradation in the low frequency sensitivity due to the quantum radiation pressure noise that results from higher levels of squeezing and/or increased laser power in the interferometers, a technique known as ``\fd\  squeezing''
\cite{KLMTV,DnD} is required ~\cite{realistic_filter_cavities,prospects_for_doubling}. In this work, we present a measurement of \fd\  squeezing from a system scalable to Advanced LIGO's upcoming A+ upgrade. To achieve this, an optical cavity with low optical losses realized a squeezing rotation frequency nearly two orders of magnitude lower than the previous demonstration~\cite{Oelker2015}. We also introduce a novel optical sensing scheme that allows for controlling the squeezed vacuum source to meet the stringent requirements of the Advanced LIGO detectors~\cite{AplusDesign}. 
The importance of \fd\  squeezing for gravitational-wave detection cannot be overstated. A+ is the only planned upgrade for Advanced LIGO in the foreseeable future, and \fd\  squeezing the only technology mature enough to provide broadband improvement in sensitivity. 

Concurrent with, and independent of, this work, a similar experiment was carried out in Japan with a \SI{300}{m} filter cavity, which demonstrates sub-shot-noise performance below the squeezed state rotation frequency \cite{TAMAFDS}.

\section{\label{sec:FDS}\Fd\  squeezing with a filter cavity}

\autofiguresvg*{
   folder=./figures/, 
   file=fc_diagram_v5a, 
   width=.9\linewidth,
   label=layout, 
   caption={
    Schematic overview of the optical and electronic layout for the experiment.
Panel A): The output of a \SI{1064}{nm} laser is used to produce a \SI{532}{nm}
field via a second harmonic generator (SHG), a local oscillator (LO) for
homodyne detection, and two frequency-shifted fields (CLF and RLF) for
generating error signals to control the squeeze angle and the filter cavity
detuning. The squeezing angle is sensed using the coherent locking field (CLF),
and the filter cavity detuning is sensed by the resonant locking field (RLF).
The \SI{532}{nm} field is split into two components, one for generating
squeezing as well as controlling the length of the OPO, and the other for controlling
the length of the filter cavity. Panel B): \fid\ squeezed vacuum generated by
the OPO is injected into the filter cavity, and experiences \fd\ rotation upon
reflection. panel C): the \fd\ squeezing is measured on a balanced homodyne
detector. Panel D): how the filter cavity will integrate optically with the LIGO
interferometer.
 }
}

\psub{How a filter cavity works} We produce \fd\  squeezing by
reflecting a squeezed state from one or more optical cavities, known as ``filter
cavities''; a method first suggested by Kimble et al. in 2001~\cite{KLMTV}. When
the filter cavity resonance is ``detuned'' with an offset relative to
the carrier frequency of the squeezed state, it applies a differential phase
shift to the upper and lower sidebands of the squeezed optical field. Since the
squeezed quadrature angle at a given sideband frequency is determined by the
relative phase between the upper and lower sidebands, the filter cavity serves
to rotate the squeezed state for sideband frequencies that lie within the cavity
linewidth (see Appendix A.3 of \cite{DnD} for details).

\psub{FD squeezing in A+} For Advanced LIGO, a single filter cavity is needed to
imprint the required frequency-dependence on the injected squeezing~\cite{realistic_filter_cavities, PhysRevD.81.122002}.
When squeezing is injected into the interferometer without
a filter cavity, as is routinely done in both Advanced LIGO \cite{O3squeezing}
and Advanced Virgo \cite{VirgoSqueezing} during observing run 3, the
optomechanical interaction between the amplitude of the optical field and the
momentum of the suspended optics will \emph{shear} (both squeeze and rotate) the
injected state at low frequencies. When injecting \fid\ 
squeezing, this rotates
part of the antisqueezed vacuum noise into the
readout quadrature~\cite{Barsotti_2018}. The A+ filter cavity is designed to apply a
compensating rotation to the squeezed state before it enters the interferometer,
to cancel the rotation due to the optomechanical interaction.

\psub{High power sets requirement on filter cavity} The power circulating in the
interferometer arm cavities determines the strength of the optomechanical
interaction and thus the frequency of the squeeze angle rotation required to
compensate it. With up to 800~kW of circulating power anticipated for
future observation runs, the filter cavity must have a bandwidth of
approximately \SI{50}{Hz}, equivalent to a storage time of
\SI{3}{ms}.

\psub{Defining cavity bandwidth} In general, for a low-loss optical resonator of length $L$ and input mirror transmission
$T_{in}$, the cavity half-width-half-maximum-power
bandwidth $\gamma/2\pi$, in units of Hz, is
\begin{equation}
  \label{eq:gammaFC}
\frac{\gamma}{2\pi} = \frac{T_{in} + \Lambda}{4\pi} \frac{c}{2L} \approx \frac{c}{4L\mathcal{F}},
\end{equation} where $\Lambda$ is the round trip loss in the cavity and
$\mathcal{F}$ is the finesse of the cavity.

\psub{Quantifying cavity bandwidth requirement} From the approximation on the
right, it follows that in order to achieve a bandwidth as narrow as \SI{50}
{Hz}, the filter cavity must either be very long or have a very high finesse
(i.e. very low optical loss). For example, a 300 m long filter cavity like the
one planned for A+ \cite{AplusDesign} (and the one operating in
Japan~\cite{TAMAloss} \cite{TAMAFDS}) can achieve a \SI{50}
{Hz} bandwidth with a finesse of $\mathcal{F} = 5,000$, while a 15 m filter
cavity would need a finesse 20 times higher, $\mathcal{F} = 100,000$.


\begin{figure*}[t]
\centering
\includegraphics[width=0.95\textwidth]{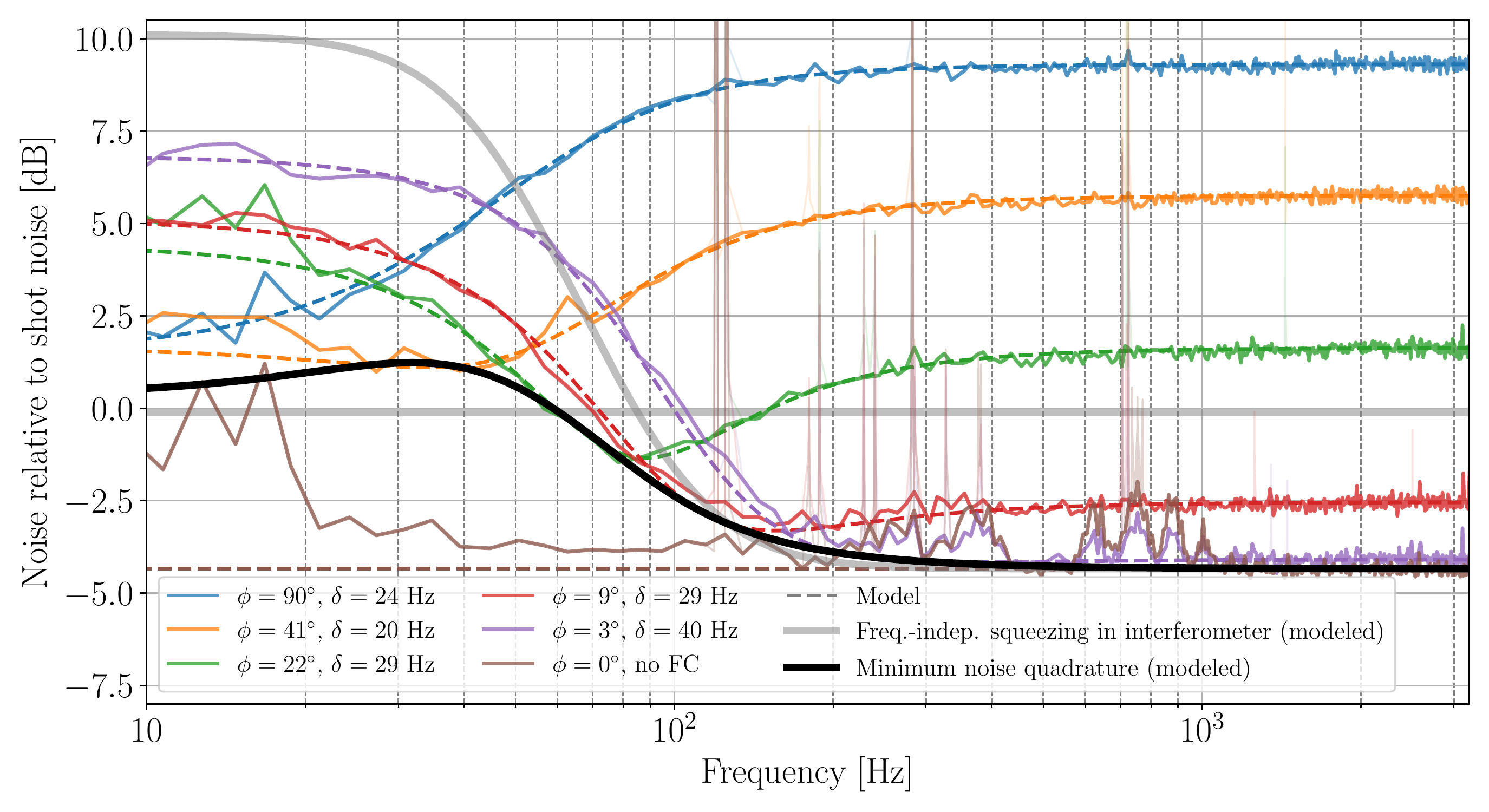}
\caption{\Fd\  squeezing at gravitational-wave detector frequencies. Measured noise (solid) is plotted alongside models (dashed).
Shot noise (gray) is shown to give a baseline of unsqueezed vacuum fluctuations.
\Fid\  squeezing (brown) shows the performance of our squeezer to low frequencies.
The action of the cavity on the squeezed vacuum is demonstrated in both squeezed (purple) and anti-squeezed (blue) quadratures, as well as intermediate homodyne angles $\phi$ (orange, green, red).
The cavity detunings $\delta$ were selected to be appropriate for
gravitational-wave detectors (20-40 Hz).
\review{ The detuning variations are from a nonlinear relationship between the
  squeezing angle and operating point of the RLF control scheme. 
  The black line shows the minimum relative quantum noise possible in an interferometer
  by squeezing after reflection by this cavity (detuned \SI{30}{Hz}).
}
\review{For a relative comparison, the gray curve models the quantum noise
  change expected from injecting only the frequency independent source of this
  experiment into a matched interferometer\cite{LigoQRPN}.
}
Data coinciding with acoustic peaks have been excluded from the frequency-binned data, but are presented in the faded traces. 
The turn-up in each curve starting at \SI{20}{Hz} is due to a mechanical
resonance of the optical table.
}
\label{fig:results}
\end{figure*}

\section{\label{sec:setup}Experimental setup}

\psub{Experimental filter cavity}
An overview of the experimental apparatus is
shown in Fig. \ref{fig:layout}. A 16 m long filter cavity is enclosed in
ultra-high vacuum chambers identical to those at the LIGO sites. The input and
output mirrors of the filter cavity are 2" super-polished fused-silica optics
mounted on tip-tilt suspensions~\cite{BramTipTilt} that are placed on
seismically-isolated platforms that mimic those in Advanced
LIGO~\cite{Matichard_2015}. The filter cavity storage time is
\SI{2.8}{\milli\second} and has a finesse of $\sim$80000 for \SI{1064}{nm}
light. The inferred cavity round-trip loss, excluding input mirror
transmissivity, is $\Lambda=19$~ppm, corresponding to a decoherence time of
\SI{5.7}{ms}~\cite{Isogai13}.


\psub{Other optical components of the experiment} The
\emph{\fid} squeezed vacuum source which drives the filter
cavity is nearly identical to the one currently in use in Advanced LIGO
\cite{O3squeezing}. Squeezing is produced by parametric downconversion in an
in-vacuum bowtie optical parametric oscillator (OPO) pumped by \SI{532}{nm}
light that is delivered via an optical fiber. The squeezed beam is then
reflected off the detuned filter cavity. A Faraday isolator steers the returning
squeezed beam from the filter cavity through a viewport towards an in-air
balanced homodyne detector~\cite{Stefszky2012}, where the \fd\ 
squeezed state is characterized. The key parameters of this system are listed in
Table \ref{tab:param}.

\begin{table}[]
\centering
\caption{Experimentally determined parameters of the \fd\  squeezed vacuum source. Entries marked by an asterisk were determined most accurately through fitting to the data. In all cases fitting produced values in agreement with independent measurements and their uncertainties.}
\label{tab:param}
\begin{tabular}{|c|c|}
  \hline
  Parameter                          & Value \\
  \hline                                                   
  \hline
    Filter cavity length               & \SI{16.0611 \pm 0.0002}{m} \\ 
    Filter cavity storage time      & \SI{2.8 \pm 0.1}{ms}\\
  Filter cavity decoherence time     & \SI{5.7 \pm 0.3}{ms}\\
  OPO nonlinear gain             & \SI{4.5 \pm  0.1}{}     \\
  OPO escape efficiency              & \SI{98 \pm 1}{\percent}   \\
  Propagation loss$^*$               & \SI{17 \pm 1}{\percent}    \\
  Homodyne visibility                & \SI{91.9 \pm 0.4}{\percent}                   \\
  Photodiode quantum efficiency      & \SI{99 \pm 1}{\percent}         \\
  Filter cavity round-trip loss       & \SI{19 \pm 1}{ppm}                   \\
 Freq. indep. phase noise (RMS) & \SI{10 \pm 5}{mrad}   \\
 Filter cavity length noise (RMS)$^*$ & \review{ $0.7(1)$ pm}     \\
Filter cavity mode matching        & \SI{92 \pm 1}{\percent}      \\

  \hline
\end{tabular}
\end{table}

\section{\label{sec:obs}Observation of \FD\  Squeezing}

\psub{Shot noise and FID squeezing measurement} Fig.~\ref{fig:results}
shows
squeezing measurements for various configurations of the filter cavity. First, a
reference measurement without squeezing is taken to determine the shot noise.
All subsequent measurements are normalized to this shot noise level. Next, a
measurement of \fid\  squeezing is taken by holding the filter
cavity far from resonance (brown). The measured squeezing level of
\SI{4.4}{dB} establishes the parameters of the squeezed vacuum
source, independent of the filter cavity.

\psub{FD squeezing measurements}
All measurements of \fd\ squeezing shown are performed
with a \SI{\sim 30} {Hz} filter cavity detuning.
With this detuning,
measurements are taken at five different homodyne angles: one for measuring the
squeezed quadrature at frequencies outside the filter cavity bandwidth (purple),
 one for measuring the antisqueezed quadrature (blue) and intermediate homodyne angles (orange, green, red).

\psub{Model fits to data} We use a detailed quantum noise model to verify our understanding of the system
and the measured parameters. In particular, we employ the two-photon
formalism, introduced in~\cite{quantum_noise_buonanno_chen} and expanded upon
in~\cite{DnD}, to model the action of the cavity on the squeezed state. Decoherence and degradation mechanisms arising from experimental
imperfections are also included. For each measurement, parameters that are difficult to measure directly (marked with asterisks in Table~\ref{tab:param}), are determined from the model fits.

\psub{FD squeezing with interferometer model} Under normal operating conditions in Advanced LIGO, the filter cavity will
rotate the squeezed state to compensate the optomechanical interaction in the interferometer~\cite{LigoQRPN}.
While the current measurement does not include the LIGO interferometer, our
model allows us to compute the maximum quantum noise reduction that could be measured 
if the interferometer's optomechanical interaction were present.
This is shown in the black trace in Fig.~\ref{fig:results}. At
frequencies near the rotation frequency (e.g., \SI{30}{Hz}), the loss in the
cavity causes this model projection to go above the shot-noise level.
This is
expected, given the finesse and optical losses of the \SI{16}{m} cavity~\cite{realistic_filter_cavities},
 and is part of the motivation for the \SI{300}{m} long
filter cavity for A+.
Optics of similar quality in a
\SI{300}{m} filter cavity will result in little degradation of squeezing even at
frequencies near the filter cavity resonance~\cite{AplusDesign}.

\section{\label{sec:control}Control of \FD\  Squeezing}

\psub{Overview of controls}
For squeezed states to be employed in gravitational-wave detectors, the phase of
the squeezed states (i.e., their orientation relative to the readout quadrature)
must be kept under tight control. Squeezed-vacuum states present a particular
challenge, as they lack a co-propagating carrier field needed for phase
measurements. The problem compounds with \fd\  squeezing, in which
the cavity rotation frequency must also be controlled.
 
\psub{Overall phase and the CLF}
As with the \emph{\fid} squeezed light sources~\cite{O3squeezing}
currently used in Advanced LIGO,
overall phase control is performed with a coherent
locking field (CLF), which is injected into the OPO and co-propagates with the
squeezed state~\cite{Vahlbruch2006a}. The CLF is offset in frequency (typically
a few MHz) relative to the carrier, in order to minimize its noise impact while
allowing a high-precision measurement of the propagation phase of the squeezed
state.

\psub{Overview of requirements} 
Deviations of the filter cavity from the nominal detuning (i.e., length offset
relative to the carrier frequency) manifest in two ways.  First, the RMS cavity
length noise causes variations in the squeezed state rotation near and below the
rotation frequency, thereby allowing some of the antisqueezed quadrature to
contaminate the readout. This is analogous to phase noise in \fid\ 
squeezing, but degrades squeezing performance at frequencies rotated by
the filter cavity. For A+, the RMS displacement noise for
the filter cavity must be lower than \SI{1}{pm}.
  
The second manifestation of filter cavity detuning noise is the
modulation of any carrier field which leaks from the
interferometer's readout port into the squeezed light path.
This leaked ``backscatter'' light undermines the benefits of squeezing by adding
noise to the readout in the gravitational-wave band~\cite{aLIGOdesign,
chua2014}. Optical isolation elements are used to reduce this leakage light, but at the
cost of increased optical loss that degrades squeezing. For A+, the
filter cavity length noise in the $\rm 10 - 100\,Hz$ band must be better than
$\rm 10^{-16}\,m/\sqrt{Hz}$.

\psub{Rotation and the filter cavity} Maintaining the rotation frequency of our
\fd\  squeezing amounts to sensing and controlling the filter
cavity length. Historically this has been done with a second field of a
different wavelength (e.g., \SI{532}{nm}) which can be introduced into the
squeezed path with dichroic optics~\cite{Oelker2015}. We use this approach to
gain initial control of the filter cavity resonance, but we find it insufficient
for high-precision control, as the cavity response has subtle discrepancies
between the two wavelengths.
To meet the stringent noise requirements of Advanced LIGO and future detectors,
we have developed a novel sensing scheme that we present here.

\psub{how RLF works}
\review{
The filter cavity sensing scheme demonstrated in this experiment adapts
the LIGO coherent control scheme \cite{O3squeezing} to prepare an
additional \SI{1064}{nm} field, frequency-shifted with respect to the
main interferometer carrier field so as to resonate in the filter cavity.
This resonant locking field (RLF) is then sensitive to the filter cavity length,
and phase or frequency noise on it will be interpreted as length noise.
The RLF is generated together with the CLF via an
acoustic-optic modulator (see Panel A in Fig. 1), and both fields
co-propagate with the squeezed vacuum field.  Co-propagation is critical
to avoid phase noise relative to the CLF, which is phase locked to the
interferometer carrier field. We estimate the RLF scheme to suppress by
at least two orders of magnitude the phase modulations arising from seismic
and acoustic noise in alternative schemes that use a \SI{532}{nm} or
\SI{1064}{nm} beam which do not benefit from the CLF phase control.
In summary, the RLF scheme has two benefits: first and foremost,
its phase coherence with the CLF and thus the squeezed state allow it to
meet the stringent backscatter requirements for A+.  Second, it
senses the cavity length at the same wavelength as the squeezing field,
thereby avoiding drifts and noises found in schemes which employ other
wavelengths for filter cavity control.
}

\section{\label{sec:future}Conclusions and Outlook}

The A+ upgrade to the Advanced LIGO detectors relies on \fd\ squeezing to
achieve a broadband reduction of quantum noise. Achieving the requisite
${<}\SI{40}{Hz}$ rotation frequency of the squeezed state necessitates a 300 m
long filter cavity with low loss optics. Here we produce this rotation frequency
by coupling the Advanced LIGO squeezed vacuum source with a 16 m filter cavity,
the longest available in our facility. The present result improves on a previous
experiment by almost two orders of magnitude~\cite{oelker2016}, providing
assurance that the 300 m long filter cavity currently under construction for A+
will work as intended. Just as important, we demonstrate the ability to control
the filter cavity length with a novel technique capable of meeting stringent
requirements imposed by back-scattered light noise. When combined with improved
optical mirror coatings, low loss Faraday isolators~\cite{Genin:18} and active
mode matching elements, \fd\ squeezing enables a broadband improvement of a
factor of 2 in the detector noise over Advanced LIGO~\cite{Aplus}, \review{
a factor of 8 in detection volume, to vastly expand LIGO's discovery space
~\cite{observingScenarios}}.

\section{Acknowledgments}
 LIGO was constructed by the California Institute
 of Technology and Massachusetts Institute of Technology with funding from the
 National Science Foundation, and operates under Cooperative Agreement No.
 PHY-0757058. Advanced LIGO was built under Grant No. PHY-0823459. The authors
 gratefully acknowledge the National Science Foundation Graduate Research
 Fellowship under Grant No. 1122374. Several people have contributed to the experiment over the years, in particular: John Miller, Georgia Mansell, Evan Hall,
 Naoki Aritomi and Vivishek Sudhir. Measurements of the optical loss of the filter cavity optics have been made by Joshua Smith, Adrian Avila-Alvarez and Juan A. Rocha at California State University, Fullerton. We also acknowledge fruitful discussions with members of the LIGO Scientific Collaboration, Virgo and KAGRA collaboration.

\bibliographystyle{apsrev4-1} 
\bibliography{references,refs_QRPN}


\end{document}

%% file: commands.tex
\def\fd{frequency-dependent}
\def\Fd{Frequency-dependent}
\def\FD{Frequency-Dependent}
\def\fid{frequency-independent}
\def\Fid{Frequency-independent}

\newcommand{\review}[1]{\textcolor{magenta}{#1}}

\newcommand{\psub}[1]{{\bf #1:}\\}

\renewcommand{\psub}[1]{}   
\renewcommand{\review}[1]{#1}  
